\documentclass[final,5p,times,twocolumn]{elsarticle}

\usepackage{amssymb}
\usepackage{lipsum}
\usepackage[version=4,arrows=pgf]{mhchem}
\usepackage[capitalise]{cleveref} 
\usepackage{siunitx} 
\DeclareSIUnit\um{\micro\m}
\usepackage{multirow} 
\usepackage{graphicx}
\usepackage{dcolumn}
\usepackage{bm}
\usepackage{booktabs}

\journal{Physics Letters B}

\begin{document}

\begin{frontmatter}

\title{Short-range correlated pairs from nucleon density profiles as a new probe of the nuclear equation of state}

\author[1]{{L. Ponnath}\corref{cor1}}
\ead{Lukas.Ponnath@york.ac.uk}
\author[2]{N. Barnea}
\author[3]{I. Korover}
\author[1]{S. Paschalis}
\author[1]{M. Petri}
\author[3]{E. Piasetzky}
\author[4,5,6,7]{X. Roca-Maza}
\author[8,9,10]{S. Typel}
\author[3]{I. Wischnevsky Shlush}

\affiliation[1]{organization={School of Physics, Engineering and Technology, University of York},
            postcode={YO10 5DD},
            city={York},
            country={United Kingdom}
}
\affiliation[2]{organization={The Racah Institute of Physics, The Hebrew University},
            postcode={9190401},
            city={Jerusalem},
            country={Israel}
}
\affiliation[3]{organization={School of Physics and Astronomy, Tel Aviv University},
            postcode={6997801},
            city={Tel Aviv},
            country={Israel}
}
\affiliation[4]{organization={Departament de Fisica Quantica i Astrofisica, Marti i Franques, 1},
            postcode={08028},
            city={Barcelona},
            country={Spain}
}
\affiliation[5]{organization={Institut de Ciencies del Cosmos, Universitat de Barcelona, Marti i Franques, 1},
            postcode={08028},
            city={Barcelona},
            country={Spain}
}
\affiliation[6]{organization={Dipartimento di Fisica ``Aldo Pontremoli'', Universit\`a degli Studi di Milano},
            postcode={20133},
            city={Milano},
            country={Italy}
}
\affiliation[7]{organization={INFN, Sezione di Milano},
            postcode={20133},
            city={Milano},
            country={Italy}
}
\affiliation[8]{organization={Technische Universität Darmstadt, Fachbereich Physik, Institut für Kernphysik},
            postcode={64289},
            city={Darmstadt},
            country={Germany}
}
\affiliation[9]{organization={GSI Helmholtzzentrum für Schwerionenforschung},
            addressline={Planckstraße 1},
            postcode={64291},
            city={Darmstadt},
            country={Germany}
}
\affiliation[10]{organization={Helmholtz Forschungsakademie Hessen für FAIR (HFHF), GSI Helmholtzzentrum für Schwerionenforschung, Campus Darmstadt},
            postcode={64289},
            city={Darmstadt},
            country={Germany}
}
\cortext[cor1]{Corresponding author}
\begin{abstract}

We investigate to what extent the observed nuclear systematics of Short-Range Correlations (SRCs) can be described by the geometry of the underlying proton and neutron density distributions. Building on previous connections between nuclear SRC contacts and one-body densities, we formulate an explicit density-overlap representation in which proton--proton, proton--neutron, and neutron--neutron SRC source terms are constructed from the full spatial density profiles obtained with Energy Density Functionals (EDFs). The framework contains two global parameters describing the overall SRC pair formation strength and the relative contribution of the spin-singlet channel, while the nucleus-dependent evolution is generated by the density-overlap integrals. The framework simultaneously reproduces several independent experimental SRC observables, including proton--proton to proton--neutron pair ratios, relative SRC pair abundances, and proton and neutron high-momentum double ratios. A comparison between relativistic and Skyrme-type EDF families demonstrates the sensitivity of these observables to the underlying density geometry and allows the effect of the complete density profiles to be distinguished from simple radius-based geometrical estimates. Applied to neutron-rich oxygen isotopes, the framework predicts a pronounced evolution of the neutron--neutron pair abundance. Comparison with an independent calculation based on occupied harmonic-oscillator wave functions and an explicit finite-distance criterion supports the predicted isotopic evolution in neutron-rich oxygen isotopes. This agreement identifies the proton--proton to neutron--neutron SRC pair ratio as a promising experimental probe of neutron-skin thicknesses in neutron-rich nuclei and, through their connection to the symmetry energy, of the nuclear equation of state.

\end{abstract}


\end{frontmatter}

\section{Introduction}
\label{introduction}
Short-Range Correlations (SRCs) constitute one of the most prominent manifestations of nuclear many-body dynamics beyond the independent-particle picture. They generate the universal high-momentum tail of nuclear momentum distributions and have been established experimentally through inclusive and exclusive electron-scattering measurements over the last two decades (see Ref.~\cite{frankfurt1993evidence, egiyan2003observation, clas2018probing, duer2019direct, hen2014momentum, schmidt2020probing, subedi2008probing} and references therein). These studies demonstrated the dominance of proton--neutron SRC pairs at intermediate relative momenta and established a remarkably universal picture of correlated nucleon pairs across the nuclear chart.

While the microscopic dynamics inside SRC pairs is now relatively well understood, considerably less is known about the geometrical conditions governing where SRC pairs preferentially form inside finite nuclei. In particular, it remains an open question to what extent the observed mass and isospin dependence of SRC observables is already determined by the underlying proton and neutron density distributions, and how much originates from genuine many-body dynamics beyond the one-body density.

Previous work has established that a substantial part of the nuclear dependence of SRC contacts \cite{WBB15} can be understood from one-body nuclear structure. Weiss et al. \cite{weiss2019short} related the proton–proton contact to the one-body proton density by constructing an uncorrelated pair density and matching it to the short-distance contact form. In a simplified uniform-density picture, this leads naturally to the approximate scaling $C_{pp}^{0}\propto Z^{2}/A$. More recently, Yankovich et al. \cite{yankovich2025relative,YPB26} generalized this geometrical interpretation to proton–proton, proton–neutron, and neutron–neutron contacts, and to the study of 3-body contacts. Within a generalized Levinger picture, the contacts were related to the one-body nuclear densities and approximated by the characteristic nuclear volumes, yielding simplified expressions governed by Z, N, and the corresponding proton, neutron, and matter radii. This demonstrated that the evolution of SRC contacts, in particular in asymmetric nuclei, can reflect differences between proton and neutron spatial distributions. This picture was further studied and verified by Liang et al. \cite{Liang24,Liang25} within the framework of the Skyrme Hartree-Fock-Bogolyubov model.

Building on this picture, we formulate an explicit density-overlap representation in which the full spatial proton and neutron density profiles enter directly through overlap integrals. Rather than reducing the nuclear geometry to characteristic radii or volumes, we use density distributions obtained from different Energy Density Functional (EDF) families. This creates a simplified framework that can be utilized to address several classes of experimental SRC observables and, importantly, neutron-rich isotopic chains where we observe that the spatial redistribution of excess neutrons produces a pronounced effect on the \textit{nn} correlations.

We find that this purely geometrical description already reproduces the dominant experimental systematics of several independent SRC observables without introducing nucleus-dependent parameters. In this work, we compare the relativistic DD2 \cite{Typel2014tqa} and Skyrme-type SAMi \cite{roca2012new} parametrization families. Within DD2, the isovector interaction is systematically varied to span symmetry-energy slope parameters from 25 -- 100 MeV in steps of 15 MeV, whereas the SAMi family spans symmetry-energy values at saturation density from 27 -- 35 MeV in steps of 1 MeV while maintaining the isoscalar properties of the functional in the refitting of the isovector parameters \cite{roca-maza2013}. Together, the two families probe a broad range of isovector density dependences within two distinct EDF formulations. A comparison between both families reveals differences in the predictions for normalized observables. These differences originate predominantly from the spatial description of the reference nucleus $^{12}$C and demonstrate that SRC observables are directly sensitive to the absolute proton and neutron density distributions, providing a natural explanation for their sensitivity to the neutron-skin thickness. Although some effects of short-range correlation physics may be effectively absorbed into EDF parameters through their calibration to experimental bulk properties, standard EDF descriptions do not explicitly reproduce characteristic SRC signatures such as high-momentum tails or the depletion of single-particle occupation numbers \cite{atar2018quasifree,paschalis2020nucleon,macchiavelli2025some, aumann2021quenching, kay2013quenching}. In the present approach, EDF calculations are therefore not used to describe the correlated high-momentum component itself, but to provide the proton and neutron density profiles that determine the geometrical overlap available for short-range pair formation. This allows us to use the spatial information contained in one-body densities to describe systematic trends in SRC pair abundances across nuclei and to introduce the $N_{pp}^{\rm SRC}/N_{nn}^{\rm SRC}$ pair ratio as a new neutron-skin-sensitive quantity with potential as a probe of the nuclear equation of state.

\section{Density-overlap framework}

Motivated by the geometrical connection between nuclear contacts and one-body densities established in Refs. \cite{weiss2019short,yankovich2025relative}, we represent the nucleus-dependent source of SRC pairs through the local overlap of the corresponding proton and neutron density distributions. Whereas the simplified geometrical model of Ref. \cite{yankovich2025relative} expresses the contacts in terms of average densities and characteristic nuclear volumes, here we retain the full radial dependence of the one-body densities.
For proton and neutron densities, $\rho_p(r)$ and $\rho_n(r)$, normalized according to
\begin{equation}
\int \rho_p(\mathbf{r})\, d^3\mathbf{r} = Z,
\end{equation}

\begin{equation}
\int \rho_n(\mathbf{r})\, d^3\mathbf{r} = N,
\end{equation}
we define phenomenological SRC source terms as
\begin{align}
I_{pp} &= \int \rho_p^2(\mathbf{r})\, d^3\mathbf{r}, \\
I_{pn} &= \int \rho_p(\mathbf{r})\,\rho_n(\mathbf{r})\, d^3\mathbf{r}, \\
I_{nn} &= \int \rho_n^2(\mathbf{r})\, d^3\mathbf{r}.
\end{align}
For approximately uniform densities, the overlap integrals reduce to $I_{pp}\sim Z^2/V_p$ and $I_{nn}\sim N^2/V_n$, where $V_p$ and $V_n$ denote the effective spatial volumes occupied by the proton and neutron distributions, respectively. Since $V_{p,n}\propto R_{p,n}^3$, this recovers the corresponding geometrical scaling $I_{pp}\sim Z^2/R_p^3$ and $I_{nn}\sim N^2/R_n^3$ \cite{yankovich2025relative}, whereas the integral representation in Eqs.~(3)--(5) retains information on the complete density profiles and their surface evolution. The densities provide a spatially resolved measure of the geometrical pair source, retaining information on both the characteristic nuclear size and the detailed shape and surface diffuseness of the density distributions. These overlap integrals quantify the geometrical probability of bringing two nucleons into close spatial proximity and therefore define the source terms for proton--proton, proton--neutron and neutron--neutron SRC pairs. \\
The proton--neutron term receives contributions from both the isospin $T=0$ and $T=1$ channels. We associate the dominant tensor contribution with the $T=0$ channel and define the corresponding number of SRC pairs as
\begin{equation}
N_{pn}^{T=0}=A_T I_{pn},
\end{equation}
where $A_T$ denotes the overall strength of the tensor-dominated channel. The $T=1$ contribution is parameterized by the relative strength $\kappa$,
\begin{equation}
N_{pn}^{T=1}=A_T\kappa I_{pn}.
\end{equation}
The total number of proton--neutron SRC pairs is therefore
\begin{equation}
N_{pn}^{\mathrm{SRC}}
=
N_{pn}^{T=0}+N_{pn}^{T=1}
=
A_T(1+\kappa)I_{pn}.
\end{equation}
Since proton--proton and neutron--neutron pairs necessarily belong to the isospin $T=1$ channel, their corresponding pair numbers are
\begin{align}
N_{pp}^{\mathrm{SRC}} &= A_T\kappa I_{pp},\\
N_{nn}^{\mathrm{SRC}} &= A_T\kappa I_{nn}.
\end{align}
Consistent with the decomposition of the one-body momentum distributions into SRC pair channels by Weiss et al. \cite{weiss2018nuclear}, each $pp$ ($nn$) pair contributes two protons (neutrons), whereas each $pn$ pair contributes one nucleon of either species. The corresponding proton and neutron SRC populations therefore become \cite{WBB15}

\begin{align}
N_p^{\mathrm{SRC}} &=A_T([1+\kappa]I_{pn}+2\kappa I_{pp}),\\
N_n^{\mathrm{SRC}} &=A_T([1+\kappa]I_{pn}+2\kappa I_{nn}),
\end{align}
from which all observables investigated in this work are constructed.

Throughout this Letter, proton and neutron densities are obtained from relativistic DD2 and non-relativistic SAMi Energy Density Functionals. Both EDF families provide systematic variations of the isovector interaction and therefore allow us to investigate the robustness of the framework with respect to different microscopic descriptions of the nuclear density distributions.

\section{Benchmark and implications}

The two global framework parameters are determined only once using experimental SRC observables. The parameter $\kappa = 0.0320$ is fixed by the measured proton--proton to proton--neutron pair ratio~\cite{duer2019direct}, while $A_T = 3.2405$ is obtained from the measured proton and neutron high-to-low momentum double ratios~\cite{clas2018probing}. Within the present framework, the high-to-low momentum ratio for nucleon species $J=p,n$ is defined as $R_J^{\mathrm{H/L}}=\chi_J/(1-\chi_J)$, where $\chi_J=N_J^{\mathrm{SRC}}/N_J$ denotes the corresponding SRC nucleon fraction. No nucleus-dependent parameters are introduced. 

\begin{figure}[htbp]
\includegraphics[width=\columnwidth]{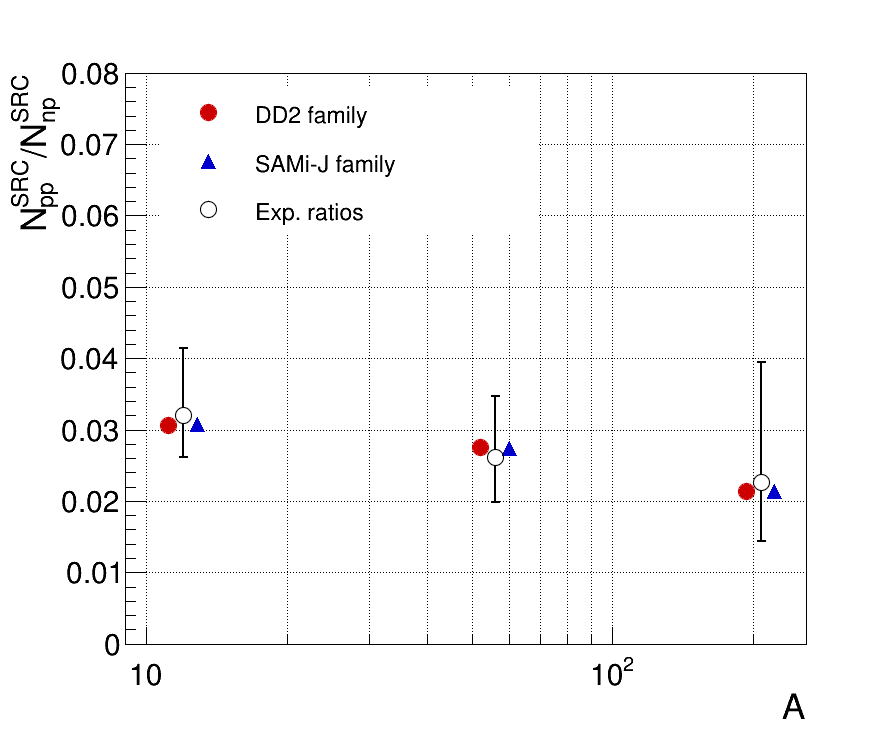}
\caption{\label{fig:pponp} Experimental and calculated $pp$ to $np$ SRC pair ratios as a function of atomic mass A. The experimental data points where taken from Ref. \cite{duer2019direct}.}
\end{figure}

We first compare the predicted proton--proton to proton--neutron SRC pair ratios with experimental data (see Figure \ref{fig:pponp}). After calibration, the framework naturally reproduces the observed weak mass dependence, indicating that the residual evolution is already contained in the density-overlap integrals.

\begin{figure}[htbp]
\includegraphics[width=\columnwidth]{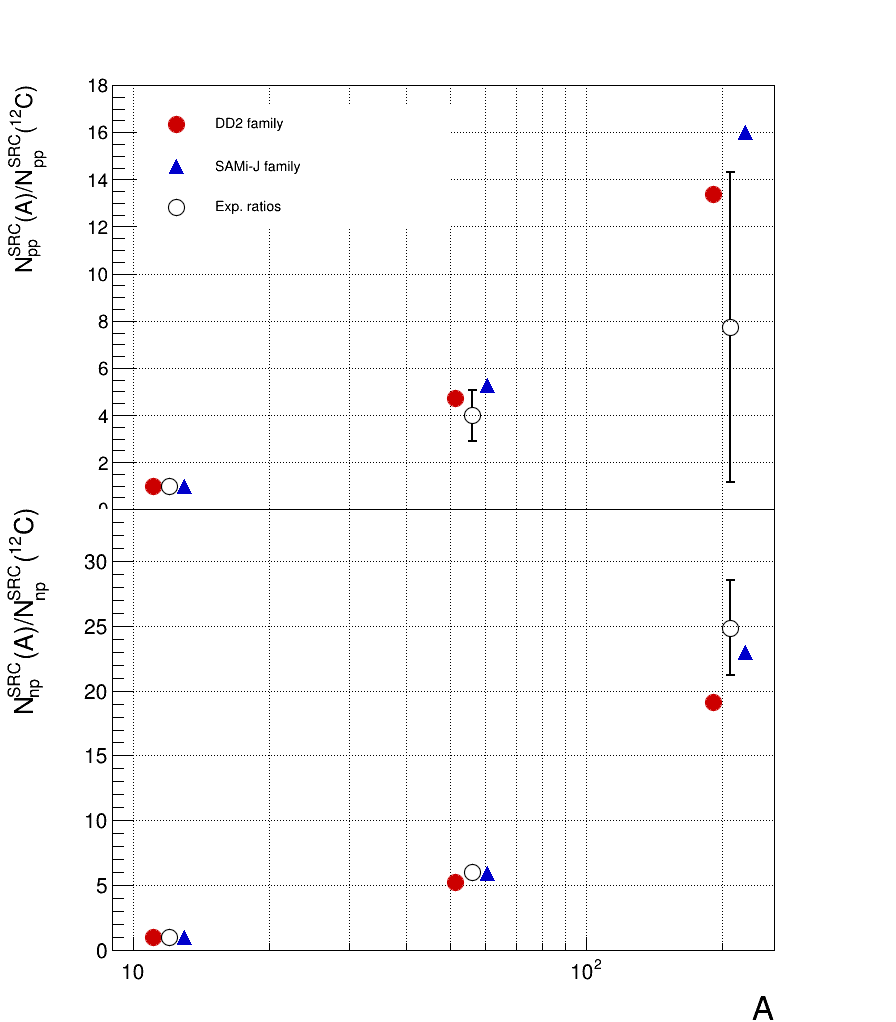}
\caption{\label{fig:relab} Number of SRC $pp$ (upper panel) and $pn$ (lower panel) pairs relative to $^{12}\mathrm{C}$ calculated within the present framework. Experimental values were taken from \cite{colle2015extracting}.}
\end{figure}

A considerably more stringent test is provided by the relative SRC abundances extracted by Colle \emph{et al.}~\cite{colle2015extracting}. These observables compare the total number of proton--neutron and proton--proton SRC pairs, as defined in Eq. (8) and (9), in heavier nuclei to the corresponding value in $^{12}\mathrm{C}$. Since the common normalization factors cancel exactly, these observables depend only on the geometrical overlap integrals. As shown in Figure \ref{fig:relab}, the calculated relative proton--proton and proton--neutron pair abundances reproduce the experimental systematics over a broad mass range without introducing additional free parameters.

\begin{figure}[htbp]
    \includegraphics[width=\columnwidth]{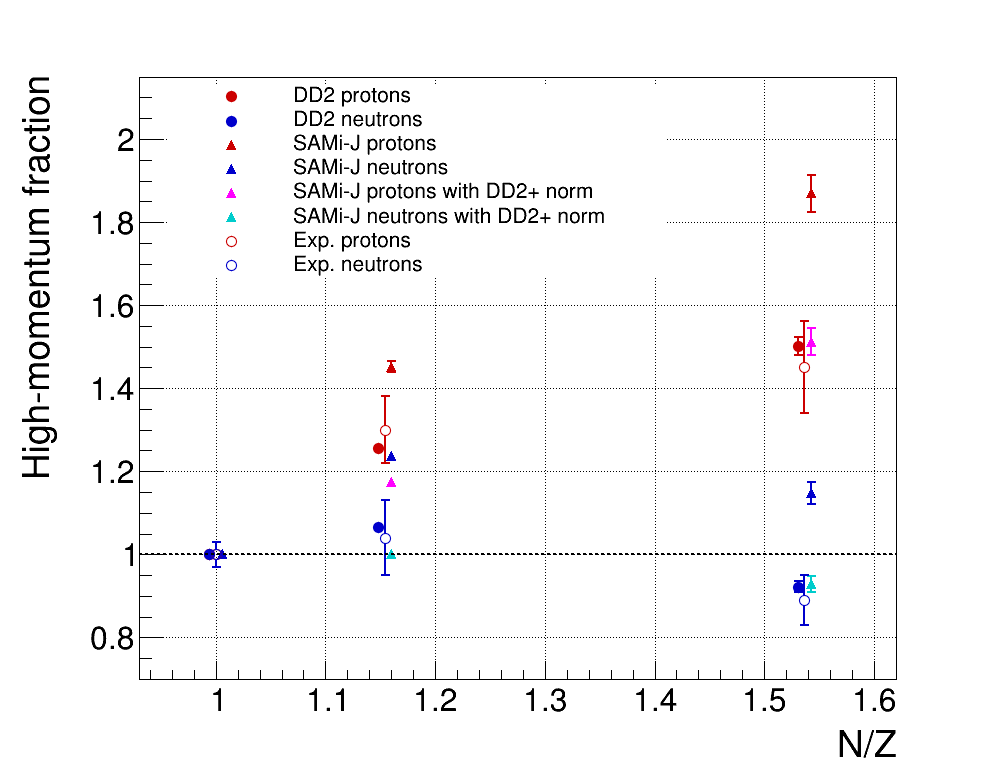}
    \caption{\label{fig:doublera}
    Double ratios of high- to low-momentum protons (red) and neutrons (blue) for $^{56}$Fe and $^{208}$Pb relative to $^{12}$C versus the corresponding neutron excess ($N/Z$). The experimental data were taken from Ref. \cite{clas2018probing}. For the filled red and blue symbols each EDF family is normalized to its own $^{12}$C prediction, corresponding to the conventional representation of the experimental double ratios. For the cyan and pink symbols the theoretical predictions of SAMi are expressed relative to a common DD2+ $^{12}$C reference in order to isolate the contribution of the normalization nucleus.}
\end{figure}

The strongest benchmark is provided by the proton and neutron high-to-low-momentum ratios relative to the corresponding value in $^{12}\mathrm{C}$ measured by Duer \emph{et al.}~\cite{clas2018probing}. The relativistic DD2 family, illustrated by the filled circles in Figure \ref{fig:doublera}, reproduces the measured trends within the experimental uncertainties. A comparison with the SAMi family (red and blue triangles) reveals a substantial discrepancy with the experimental data. This discrepancy can, however, largely be traced back to the normalization to $^{12}$C.

To isolate this effect, all relative observables were additionally normalized to the $^{12}$C prediction of the central DD2+ interaction (cyan and magenta triangles), which reproduces the experimentally measured charge radius of $^{12}$C more accurately than the SAMi family. Under this common normalization, the differences between both EDF families are significantly reduced. This demonstrates that the apparent EDF dependence of the benchmark observables is dominated by the description of the reference nucleus rather than by the heavier nuclei themselves.

This observation has an important implication. It shows that the density-overlap framework is sensitive to the absolute proton and neutron density distributions, particularly for light nuclei, rather than only to their relative evolution across isotopic chains. Consequently, nuclei with accurately known charge and neutron radii provide ideal benchmarks for validating the geometrical component of the framework. Once this geometrical input is constrained experimentally, the remaining differences can be attributed more directly to the underlying SRC dynamics.

Having established that the dominant experimental SRC systematics are already encoded in the geometrical overlap of one-body densities, the framework can now be employed to investigate new observables governed by the same underlying geometry. The observed sensitivity to the absolute proton and neutron density distributions naturally raises the question whether the same geometrical framework can be used to identify SRC observables that are directly sensitive to neutron skins. Since modern EDFs predict systematic changes of the proton and neutron density profiles with the isovector interaction, the density-overlap framework provides a direct connection between SRC pairing probability and neutron-skin-sensitive observables.

\section{A neutron-skin-sensitive SRC pair ratio}
Among the quantities naturally emerging from the present framework, the ratio between proton--proton and neutron--neutron SRC pairs is of particular interest, as it is completely independent of the global framework parameters and given by

\begin{equation}
\frac{N_{pp}^{\mathrm{SRC}}}{N_{nn}^{\mathrm{SRC}}}
=
\frac{I_{pp}}{I_{nn}},
\end{equation}

where all dependence on the overall SRC normalization and on the reduced strength of the same-isospin pairing cancels exactly. Consequently, this pair ratio is determined exclusively by the geometrical overlap of the proton and neutron density distributions and therefore provides a direct probe of the spatial structure of the nucleus.

With increasing neutron-skin thickness, the neutron density becomes spatially more extended, while the proton density changes only moderately. Since the neutron--neutron source term is proportional to $I_{nn}=\int \rho_n^2(\mathbf{r})\,d^3r, $ the redistribution of a fixed number of neutrons over a larger volume reduces the integrated neutron self-overlap. In contrast, the proton self-overlap $I_{pp}$ remains comparatively stable. Consequently, the ratio $ N_{pp}^{\mathrm{SRC}}/N_{nn}^{\mathrm{SRC}} = I_{pp}/I_{nn} $ increases with neutron-skin thickness. To quantify this behaviour, this ratio was calculated for both relativistic DD2 and Skyrme-type SAMi Energy Density Functionals spanning a broad range of isovector parameters.

\begin{figure}[htbp!]
\includegraphics[width=\columnwidth]{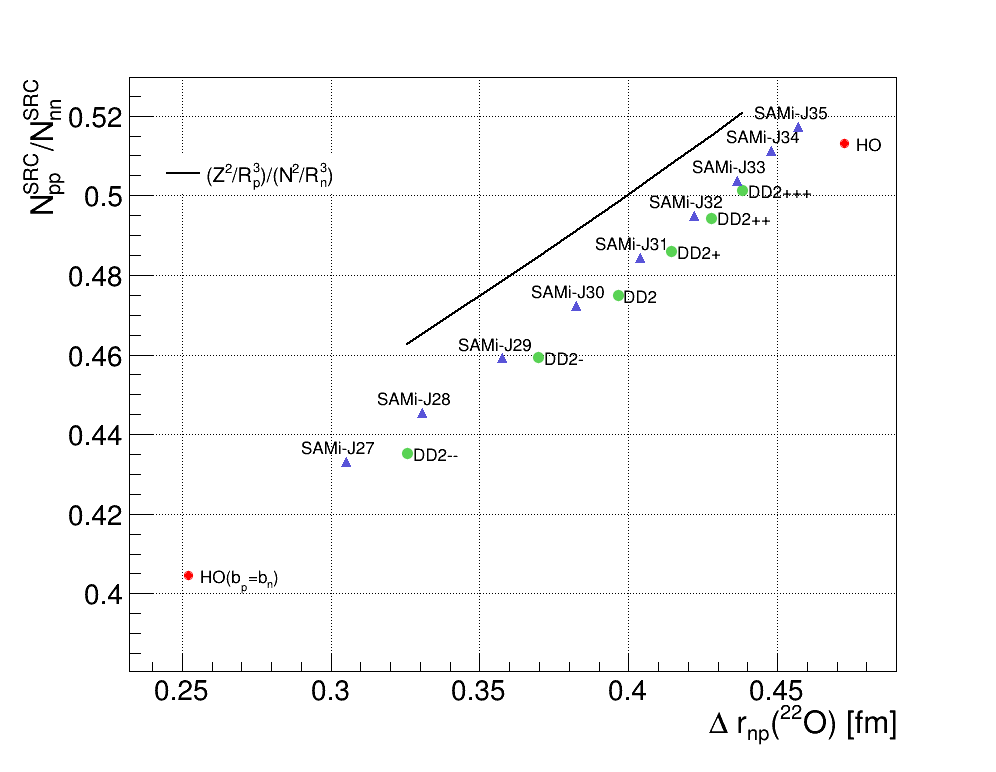}
\caption{\label{fig:O22} Ratio of $pp$ to $nn$ SRC pairs for $^{22}$O versus the corresponding neutron-skin thickness. The different data points were calculated with the relativistic DD2 interactions with the symmetry-energy slope parameter L ranging from 25 to 100 MeV, and non-relativistic SAMi interactions with symmetry energy J values ranging from 27 to 35 MeV. The HO points are calculations performed in the present work using the wave-function method introduced in Ref. \cite{wischnevsky2026modeling}. See text for details.}
\end{figure}

Figure \ref{fig:O22} shows the calculated ratio for the neutron-rich $^{22}$O isotope as a function of the corresponding neutron-skin thickness. The black line illustrates the corresponding prediction obtained from the simplified geometrical approximation $N_{pp}^{\mathrm{SRC}}/N_{nn}^{\mathrm{SRC}} \simeq (Z^2/{R_p}^3)/(N^2/{R_n}^3)$ \cite{yankovich2025relative}. 

It is instructive to compare the present density-overlap framework with the wave-function-based method introduced in Ref. \cite{wischnevsky2026modeling}. In that method, the probability for two nucleons occupying specified harmonic-oscillator (HO) shell-model orbitals to satisfy the finite-distance condition $|\mathbf{r}_1-\mathbf{r}_2| < r_{\mathrm{SRC}}$ is evaluated explicitly, including the spatial-exchange symmetry of the two-nucleon wave function. Here, we apply this previously developed method to the oxygen isotopic chain. These oxygen-isotope calculations and their comparison with the density-overlap predictions are new results of the present work. In addition, results were computed for the case of identical proton and neutron oscillator length ($\mathrm{HO(b_{p}=b_{n})}$).
The two formulations describe the geometrical conditions for SRC formation at different levels. The HO method resolves the contributions of occupied single-particle orbitals and employs an explicit finite relative-distance criterion. The density-overlap framework instead sums the single-particle information into the proton and neutron densities and represents the geometrical SRC source through $I_{ij} = \int \rho_i(\mathbf{r})\,\rho_j(\mathbf{r})\, d^3\mathbf{r}$ . In the short-range limit, the finite-range probability calculated from the occupied wave functions is, to leading order, proportional to the SRC volume multiplied by the corresponding local density-overlap integral. Thus, the two approaches share the same underlying geometrical picture, although they retain different nuclear-structure information. The HO method includes shell occupancies, orbital-pair contributions, exchange symmetry, and finite-range effects, whereas the EDF approach provides self-consistent radial density profiles, including their surface diffuseness and neutron-skin evolution.

Despite their fundamentally different microscopic formulations, both EDF families and the HO calculation follow nearly the same correlation between the SRC pair ratio and the neutron-skin thickness. The HO result extends this comparison to a somewhat larger value of the neutron-skin thickness and remains consistent with the continuation of the EDF trend. This agreement provides an independent test of the density-overlap interpretation: the correlation persists when close-proximity pairs are evaluated directly from occupied shell-model wave functions.
The radius-based estimate captures the leading dependence on neutron-skin thickness, while its deviations from the EDF and HO calculations reflect information not contained in the rms radii alone. Such effects include the detailed radial shapes and surface diffuseness of the densities, shell structure, spatial-exchange symmetry, and the finite distance used to define a close-proximity pair.

This result is consistent with the benchmark analysis presented above, which demonstrated that the framework is directly sensitive to the absolute proton and neutron density distributions. Since the neutron-skin thickness characterizes the relative spatial extension of these distributions, its correlation with the geometrical SRC source terms emerges naturally within the density-overlap picture.

Within the DD2 family, systematic variations of the isovector interaction correspond to symmetry-energy slope parameters ranging from approximately 25 to 100~MeV. Owing to the strong correlation between neutron-skin thickness and the symmetry-energy slope parameter, the predicted SRC pair ratio exhibits an approximately linear dependence on $L$ throughout this interval.
\begin{figure}[htbp!]
\includegraphics[width=\columnwidth]{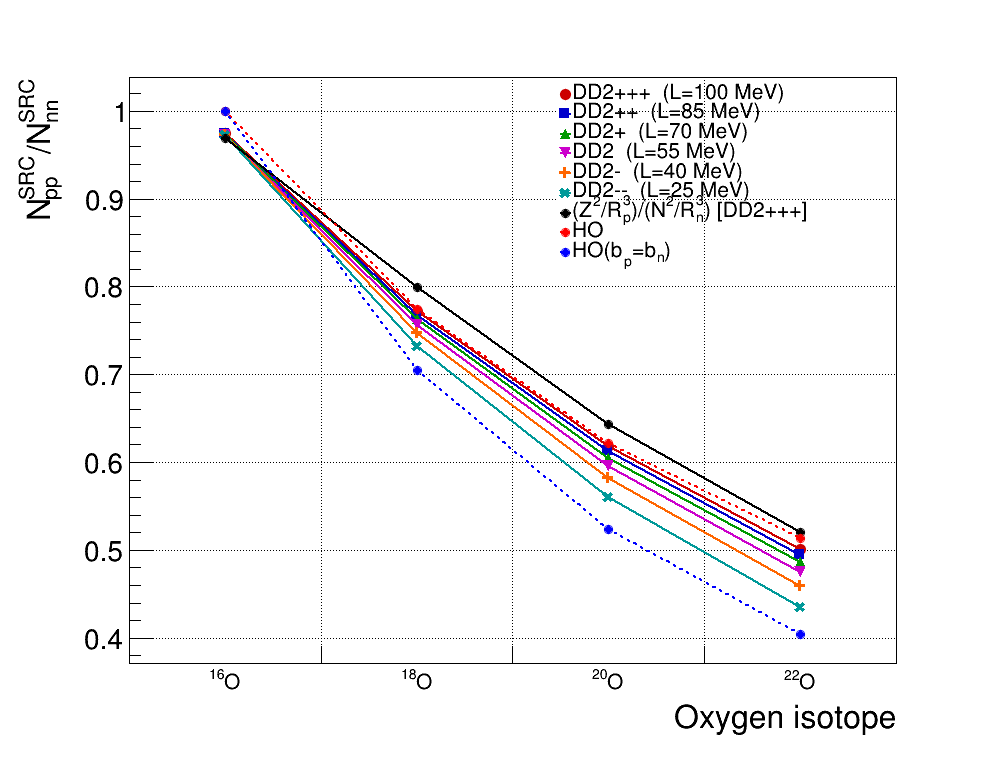}
\caption{\label{fig:Sens} Ratio of pp to nn SRC pairs for different oxygen isotopes calculated with the relativistic DD2 interaction with the symmetry-energy slope parameter ranging from 25 to 100 MeV. The HO points are calculations performed in the present work using the wave-function method introduced in Ref. \cite{wischnevsky2026modeling}. See text for details.}
\end{figure}
For an experimental application, the absolute ratio $N_{pp}^{\mathrm{SRC}}/N_{nn}^{\mathrm{SRC}}$ for a neutron-rich nucleus can be expressed relative to the symmetric reference nucleus from the same isotopic chain, thereby reducing common systematic uncertainties, such as detector efficiencies, acceptances and reaction-model uncertainties. Figure \ref{fig:Sens} shows the resulting evolution along the oxygen isotopic chain for the controlled DD2 family. The black line in Figure 5 shows the corresponding evolution obtained from the simplified radius-based estimate. 
The standard HO calculations show a similar trend as the EDF results. This agreement is nontrivial because these approaches describe nuclear geometry through self-consistent density profiles and HO shell-model orbitals, respectively. Their common isotopic dependence indicates that the leading trend is robust and is governed primarily by the changing spatial overlap of the proton and neutron distributions.
The $\mathrm{HO(b_{p}=b_{n})}$ calculation exhibits a noticeably steeper decrease toward the neutron-rich isotopes. This behavior demonstrates the sensitivity of the predicted ratio to the assumed relative radial scales of the proton and neutron wave functions and illustrates the limitations of a restricted HO radial form for describing the surface evolution of neutron-rich nuclei. The spread between the two HO prescriptions provides an indication of this model dependence. In contrast, the agreement between the standard HO and EDF calculations supports the conclusion that the dominant isotopic evolution is controlled by nuclear geometry rather than by the particular representation used to describe it.\\
Using the EDF correlation generated by the systematic variation of $L$, a relative experimental precision of approximately $2\%$ in the isotopic SRC ratio would correspond, within the present framework, to an uncertainty of about $\Delta L \simeq 11$~MeV. The oxygen isotopes shown in Figure \ref{fig:Sens} provide one representative example of such a measurement.

\section{Conclusions}

Building on earlier connections between SRC contacts and one-body nuclear densities, we have developed an explicit density-overlap representation that retains the full spatial dependence of the proton and neutron distributions. The formulation generalizes simple radius- and volume-based geometrical estimates to realistic density profiles obtained from modern Energy Density Functionals and allows the geometrical picture to be tested directly against several independent classes of experimental SRC observables.

The benchmark analysis demonstrates that the dominant nuclear dependence of experimentally accessible SRC observables is already encoded in the geometrical overlap of the underlying proton and neutron density distributions. A comparison between relativistic DD2 and Skyrme-type SAMi EDF families further shows that the benchmark observables are directly sensitive to the absolute proton and neutron density distributions and highlights the importance of accurately describing the reference nucleus when constructing relative SRC observables.

The framework provides a phenomenological separation between a universal interaction-strength component and a nucleus-dependent geometrical component. This interpretation naturally explains the observed sensitivity of the SRC source terms to neutron-skin thicknesses and leads to the prediction that the proton--proton to neutron--neutron SRC pair ratio constitutes a parameter-independent quantity with strong sensitivity to neutron skins and the symmetry-energy slope parameter.

The present work extends the established geometrical connection between one-body nuclear structure and SRCs into a spatially resolved framework applicable to neutron-rich nuclei. Its combination with systematic EDF variations identifies SRC pair ratios along isotopic chains as particularly sensitive probes of the redistribution of neutron density. The pronounced evolution predicted for the oxygen chain, especially in the $nn$ channel, provides a concrete experimental test of this picture and may offer complementary constraints on neutron skins and the density dependence of the nuclear symmetry energy.

\section*{Acknowledgements}
We acknowledge A. O. Macchiavelli for insightful discussions. This work was supported by the UK STFC under grant number ST/Y000285/1 (L.P, M.P and S.P). X.R.M. acknowledges support by MI-CIU/AEI/10.13039/501100011033 and by FEDER UE through grants PID2023-147112NB-C22 and CNS2025-165430; and through the ``Unit of Excellence Maria de Maeztu 2025-2028'' award to the Institute of Cosmos Sciences, grant CEX2024-001451-M. Additional support is provided by the Generalitat de Catalunya (AGAUR) through grant 2021SGR01095. The work of N.B. was supported by the Pazy Foundation, grant number 443/2021. The work of E.P. was supported by the Israeli Science Foundation (ISF) under Grants No. 917/20, and No. 371/25, and by the Pazy Foundation 520/23, and the ISF NSFC joint research program Grant No. 3107/23. The work of I.K. was supported by ISF 830/24, and by Pazy Foundation 737/25.
\appendix 

\bibliographystyle{ieeetr} 
\bibliography{example}

\end{document}